\documentclass[runningheads,a4paper]{llncs}

\usepackage{amssymb}
\usepackage{graphicx}
\usepackage{url}
\usepackage{amsmath}
\usepackage{algorithm}
\usepackage{algpseudocode}

\let\oldsim\sim 
\renewcommand{\sim}{{\oldsim}}

\algnewcommand{\LeftComment}[1]{\Statex \(\triangleright\) #1}

\begin{document}

\mainmatter 

\title{A Note On The Natural Range Of Unambiguous-SAT}


\author{Tayfun Pay \thanks{This work was conducted during my PhD studies at the Graduate Center of New York, intermittently between 2008 and 2012. However, it was never put into LaTex and released, until now. }}

\authorrunning{Tayfun Pay}

\urldef{\mailsa}\path|tpay@gradcenter.cuny.edu|

\institute{ {Graduate Center of New York \\ 365 5$^{th}$ Avenue,\\ New York, New York 10016\\ \mailsa}}


%
%

\maketitle

\begin{abstract}
We discuss the natural range of the Unambiguous-SAT problem with respect to the number of clauses. We prove that for a given Boolean formula in precise conjunctive normal form with $n$ variables, there exist functions $f(n)$ and $g(n)$ such that if the number of clauses is greater than $f(n)$ then the formula does not have a satisfying truth assignment and if the number of clauses is greater than $g(n)$ then the formula either has a unique satisfying truth assignment or no satisfying truth assignment. The interval between functions $f(n)$ and $g(n)$ is the \emph{natural range} of the ${Unambiguous-SAT}$ problem. We also provide several counting rules and an algorithm that determines the unsatisfiability of some formulas in polynomial time.
\end{abstract}

$\newline$
\section{Introduction}

The study of restricted non-deterministic polynomial-time Turing machines  was introduced in \cite{V76} with the complexity class {\bf UP}. This complexity class possesses what is called a promise problem rather than a complete problem. The promise problem for {\bf UP} is called ${Unambiguous-SAT}$, which asks: given a Boolean formula in Conjunctive Normal Form (CNF) that is promised to have a unique satisfying truth assignment or none, does it have a unique satisfying truth assignment? This is a totally different problem from the ${Unique-SAT}$ problem, which is a complete problem for complexity class {\bf US} that was defined in \cite{BG82}. The ${Unique-SAT}$ problem asks: given a Boolean formula in CNF, is it true that it has a unique satisfying truth assignment? The isolation lemma in \cite{VV86} showed that {\bf NP} $\subseteq$ {\bf RP }$^{Unambiguous-SAT}$, where {\bf RP} is the randomized polynomial time that was defined in \cite{G77}. The consequence of this was that if ${Unambiguous-SAT}$ could be solved in polynomial time then {\bf RP} = {\bf NP}.\footnote{More thorough explanation of complexity classes and the relationships between them can be found in \cite{HO02}.\nocite{CP18}}

In this paper, we define what we call the precise conjunctive normal form (PCNF). We show the total number of different clauses that can be obtained for Boolean formulas in PCNF with $n$ variables. We also explain a way to go from Boolean formulas in CNF to Boolean formulas in PCNF in polynomial time. We then prove two key results regarding the total number of clauses for Boolean formulas in PCNF with $n$ variables: 

1) The \underline{final point of satisfiability}, denoted by function $f(n)$, such that if there are more than $f(n)$ many clauses, then the given Boolean formula in PCNF will have no satisfying truth assignment. 

2) The \underline{last point of double satisfiability}, denoted by function $g(n)$, such that if there are more than $g(n)$ many clauses, then the given Boolean formula in PCNF will either have a unique satisfying truth assignment or no satisfying truth assignment. 

Based on these results, the \emph{natural range} of the ${Unambiguous-SAT}$ problem lies between these two points for Boolean formulas in PCNF.

We also provide several additional functions that allow us to determine the unsatisfiability of a given Boolean formula in PCNF by counting the number of literals and variables. Furthermore, we present a polynomial time algorithm that determines the unsatisfiability of some Boolean formulas in PCNF by counting the number of clauses. 
$\newline$

\section{Defining Precise Conjunctive Normal Form}

A Boolean formula consists of one or more literals $\{l_{1}, l_{2}, l_{3}, ...\}$ that are separated by conjunction ($\wedge$) and/or disjunction ($\vee$) operators. A literal can be a variable ($a$) or its complement ($\sim a$), which is obtained by using the negation ($\sim$) operator, and it can be assigned the value True or False. A Boolean formula is said to be satisfiable if some truth assignment to its variables makes the Boolean formula evaluate to True. On the other hand, a Boolean formula is said to be unsatisfiable if all possible combinations of truth assignments to its variables makes the Boolean formula evaluate to False.

\begin{definition}\normalfont
A Boolean formula in Conjunctive Normal Form (CNF) consists of conjunction of clauses, where the clauses are a disjunction of literals. 
\end{definition}

There does not appear to be a definitive definition of CNF that is universally used in the literature. There are some definitions where repetition of clauses are not allowed. Other definitions do allow repetition of clauses, but disallow repetition of variables and/or literals within a clause. And so on. The various variants of CNF that are used in the literature can be found in \cite{AMH09}. As a result, we wanted to define the most exact definition of CNF. Our definition is similar to the definition of ``random k-CNF formulas" in chapter 8 of \cite{AMH09}, with two differences. It neither requires randomness nor does it require exactly k literals per clause.

\begin{definition}\normalfont
A Boolean formula in Precise Conjunctive Normal Form (PCNF) consists of conjunction of unique clauses, where the clauses are a disjunction of unique literals and do not contain both the variable and its complement.
\end{definition}

As can be seen from the definitions, PCNF formulas are similar to CNF formulas but with subtle differences. PCNF formulas do not allow repetition of clauses within a given formula, nor do they allow repetition of variables in a given clause, whether in complemented or uncomplemented forms. For example, the following three clauses are not allowed in PCNF formulas: 
\begin{enumerate}
\item $({a}\vee {a}\vee \sim {a})$ 
\item $({a} \vee {a} \vee {b})$
\item $({a} \vee \sim{a} \vee {b})$
\end{enumerate}

However, a CNF formula that does not adhere to these rules can be easily transformed into a PCNF formula that does with a single scan of the input formula. To transform a CNF formula into a PCNF formula, iterate over all of the clauses one by one. First, remove any literals that appear more than once in each clause. Next, discard clauses that contain a variable and its complement, as they are always valid. Finally, discard any duplicate clauses.\footnote{This algorithm can be seen in the appendix section.}

\section{Finding the \emph{natural range} of ${Unambiguous-SAT}$}

In this section, we prove three theorems related to Boolean formulas in PCNF by using proof by case analysis. We also show some intriguing relationships between Pascal's triangle and the satisfiability of some Boolean formulas in PCNF with respect to their total number of clauses. In essence, we establish the final point of satisfiability and the last point of double satisfiability with respect to the total number of clauses for Boolean formulas in PCNF. 

\begin{theorem}\normalfont

A Boolean formula in Precise Conjunctive Normal Form can have up to $ m(n) = \sum_{i=1}^{n} 2^{i} \binom{n} {i}$ different clauses, where $n$ is the number of variables.
\begin{proof} Let's start by counting the total number of clauses we can have for a Boolean formula in PCNF.
\begin{itemize}
\item When n=1, there are 2 clauses in a Boolean formula in PCNF and these clauses are as follows: $(a)$ and $(\sim a)$.
 
\item When n=2, there are 8 clauses in a Boolean formula in PCNF and these clauses are as follows:
     \begin{itemize}
    \item 4 clauses with a single variable:$(a)$, $(\sim a)$, $(b)$ and $(\sim b)$.
   \item 4 clauses with double variables: $({a}\vee {b})$, $({a}\vee \sim {b})$, $(\sim {a}\vee {b})$ and $(\sim {a}\vee \sim {b})$.
     \end{itemize}
\item When n=3, there are 26 clauses in a Boolean formula in PCNF and these clauses are as follows:
    \begin{itemize}
    \item 6 clauses with a single variable: $(a)$, $(\sim a)$, $(b)$, $(\sim b)$, $(c)$ and $(\sim c)$.
    \item 12 clauses with double variables: $({a}\vee {b})$, $({a}\vee \sim {b})$, $(\sim {a}\vee {b})$,  $(\sim {a}\vee \sim {b})$, $({a}\vee {c})$, $({a}\vee \sim {c})$, $(\sim {a}\vee {c})$,  $(\sim {a}\vee \sim {c})$, $({b}\vee {c})$, $({b}\vee \sim {c})$, $(\sim {b}\vee {c})$ and $(\sim {b}\vee \sim {c})$.
    \item 8 clauses with triple variables: $({a}\vee {b} \vee {c})$, $({a}\vee {b} \vee \sim {c})$, $({a}\vee \sim {b} \vee {c})$, $({a}\vee \sim {b} \vee \sim {c})$, $(\sim {a}\vee {b} \vee {c})$, $(\sim {a}\vee {b} \vee \sim {c})$, $( \sim {a}\vee \sim {b} \vee {c})$ and $( \sim {a}\vee \sim {b} \vee \sim {c})$.
    \end{itemize}
\end{itemize}

We will not write these clauses down one by one, but the distribution of the clauses for when n=4 are as follows:
8 clauses with a single variable, 24 clauses with double variables, 32 clauses with triple variables and a 16 clauses with quadruple variables. And the distribution of the clauses for when n=5 are as follows: 10 clauses with a single variable, 40 clauses with double variables, 80 clauses with triple variables, 80 clauses with quadruple variables and 32 clauses with quintuple variables. 

It is easy to see that as n increases, a new pair of single clauses are introduced. This can be expressed with the following formula: $m_{1}(n) = 2^1 \binom{n} {1}$. The increase in the number of clauses with double variables can also be expressed with a similar formula: $m_{2}(n) = 2^2 \binom{n} {2}$. And the same thing can be said about the clauses with triple variables, where the formula is $m_{3}(n) = 2^3 \binom{n} {3}$. And so on and so forth as n increases. The exponential part of the formulas give us the total number of different variations of the selected variables in complemented and uncomplemented forms. The binomial part of the formulas give us the total number of different ways we can create clauses by selecting that many variables from the total number of variables, which is n. The binomial values in each $m_{k}$ correspond to the values at the k$^{th}$ diagonal of the Pascal's triangle starting from the right and going left as k increases, where k=0 is right most diagonal with all 1 values. \footnote{You can also clearly start from the left and go right!} The combined formula that gives the total number of possible clauses is $ m(n) = \sum_{i=1}^{n} 2^{i} \binom{n} {i}$, where n is the number of variables.

Let's assume that m(n) is correct for n. We now look at n\`{} = n+1, the successor of n. It can be seen that m(n\`{}) properly determines the total number of possible clauses in a Boolean formula in PCNF when there are n+1 variables. Thus $m(n+1) = m(n) + \sum_{i=1}^{n} 2^{i} \binom{n-1} {i-1}$, where the binomials corresponds to the values at the n$^{th}$ row of the Pascal's triangle, where n=0 is the top row with a single value of 1.

Therefore, a Boolean formula in Precise Conjunctive Normal Form can have up to $ m(n) = \sum_{i=1}^{n} 2^{i} \binom{n} {i}$ different clauses, where $n$ is the number of variables. $\blacksquare$

\end{proof}
\end{theorem}

The formula $ m(n) = \sum_{i=1}^{n} 2^{i} \binom{n} {i}$ actually simplifies to $3^{n}-1$.
\newpage

\begin{theorem}\normalfont

If a Boolean formula in Precise Conjunctive Normal Form has more than $f(n) = \sum_{i=1}^{n} 2^{i} \binom{n} {i} - \sum_{i=1}^{n} \binom{n} {i}$ clauses then it has no satisfying truth assignment.
\begin{proof} 

We would need to determine the maximum number of clauses that a Boolean formula in PCNF can have such that adding one more clause would result in a formula with no satisfying truth assignment. We can consider a Boolean formula in PCNF with all possible m(n) clauses and then make an arbitrary truth assignment to the variables. For instance, we can assign True to all variables in their uncomplemented forms.\footnote{What we exactly mean is that $a=True$, $b=True$, $c=True$, $d=True$, $e=True$, and so on and so forth for all n variables.} We can then count the total number of clauses that are satisfied by this assignment. This will give us the maximum number of clauses in a Boolean formula in PCNF such that adding one more clause would result in a formula with no satisfying truth assignment. 

\begin{itemize}
\item When n=1, there is only 1 clause in the Boolean formula in PCNF that can be satisfiable and this clause is as follows: $(a)$.
 
\item When n=2, there are 5 clauses in the Boolean formula in PCNF that can be satisifiable and these clauses are as follows:
    \begin{itemize}
    \item 2 clauses with a single variable:$(a)$ and $(b)$.
    \item 3 clauses with double variables: $({a}\vee {b})$, $({a}\vee \sim {b})$ and $(\sim {a}\vee {b})$.
     \end{itemize}
\item When n=3, there are 19 clauses in the Boolean formula in PCNF that can be satisfiable and these clauses are as follows:
    \begin{itemize}
    \item 3 clauses with a single variable: $(a)$, $(b)$ and $(c)$.
    \item 9 clauses with double variables: $({a}\vee {b})$, $({a}\vee \sim {b})$, $(\sim {a}\vee {b})$,  $({a}\vee {c})$, $({a}\vee \sim {c})$, $(\sim {a}\vee {c})$,  $({b}\vee {c})$, $({b}\vee \sim {c})$ and $(\sim {b}\vee {c})$
    \item 7 clauses with triple variables: $({a}\vee {b} \vee {c})$, $({a}\vee {b} \vee \sim {c})$, $({a}\vee \sim {b} \vee {c})$, $({a}\vee \sim {b} \vee \sim {c})$, $(\sim {a}\vee {b} \vee {c})$, $(\sim {a}\vee {b} \vee \sim {c})$ and $( \sim {a}\vee \sim {b} \vee {c})$.
    \end{itemize}
\end{itemize}

When n=4 then there are 65 clauses in the Boolean formula in PCNF that can be satisfiable and the distribution of these clauses are as follows: 4 clauses with a single variable, 18 clauses with double variables, 28 clauses with triple variables and a 15 clauses with quadruple variables. When n=5 then there are 211 clauses in the Boolean formula in PCNF that can be satisfiable and the distribution of these clauses are as follows: 5 clauses with a single variable, 30 clauses with double variables, 70 clauses with triple variables, 75 clauses with quadruple variables and 31 clauses with quintuple variables. 

It is easy to see that as n increases, we would always remove one additional clause with a single variable. This can be expressed with the following formula: $r_{1}(n) = \binom{n} {1}$. The number of clauses with double variables that we would need to remove can also be expressed with a similar formula: $r_{2}(n) = \binom{n} {2}$. And the same thing can be said about the clauses with triple variables, where the formula is $r_{3}(n) = \binom{n} {3}$. And so on and so forth as n increases. Once again, the binomial values in each $r_{k}$ correspond to the values at the k$^{th}$ diagonal of the Pascal's triangle starting from the right and going left as k increases, where k=0 is right most diagonal with all 1 values. The combined formula that gives the total number of clauses we would need to remove to achieve a satisfiable formula from a Boolean formula in PCNF with m(n) clauses is  $ r(n) = \sum_{i=1}^{n} \binom{n} {i}$. 

Let's assume that r(n) is correct for n. We now look at n\`{}= n+1, the successor of n. It can be seen that r(n\`{}) properly determines the total number of clauses we would need to remove to achieve a satisfiable formula when there are n+1 variables in a Boolean formula in PCNF. Thus $r(n+1) = r(n) + 2^{n} $, where $2^{n}$ equals to the difference in the values between each row of Pascal's triangle. 

Therefore, if a Boolean formula in Precise Conjunctive Normal Form has more than f(n) = m(n) - r(n) = $\sum_{i=1}^{n} 2^{i} \binom{n} {i} - \sum_{i=1}^{n} \binom{n} {i}$ clauses then it has no satisfying truth assignment. $\blacksquare$

\end{proof}
\end{theorem}

The formula $ f(n) =  \sum_{i=1}^{n} 2^{i} \binom{n} {i} - \sum_{i=1}^{n} \binom{n} {i}$ actually simplifies to $3^{n}-2^{n}$. This is the final point of satisfiability.

\begin{theorem}\normalfont

If a Boolean formula in Precise Conjunctive Normal Form has more than $g(n) = \sum_{i=1}^{n} 2^{i} \binom{n} {i} - \sum_{i=1}^{n} \binom{n} {i} - \sum_{i=0}^{n-1} \binom{n-1} {i} $ clauses then it has either a unique satisfying truth assignment or no satisfying truth assignment.
\begin{proof}

We can assume that we are starting with the Boolean formulas in PCNF from the previous theorem with f(n) many clauses, and that the variables in their uncomplemented forms (a, b, c, d, e, ...) are set to True. We would like to flip the truth assignment of one of the variables and then remove all clauses that become unsatisfiable as a result. This will provide us with formulas that can be satisfied by two different truth assignments. Adding one more clause would then make these formulas either satisfiable by only one truth assignment or unsatisfiable by any truth assignment. 

This approach does not apply when n=1, because a Boolean formula in PCNF with only one variable can only have either a unique satisfying truth assignment or no satisfying truth assignment. We will therefore start with Boolean formula in PCNF with two variables. Let's also assume that we flipped the truth assignment of one variable (a) from True to False to obtain the second truth assignment.\footnote{The first truth assignment is as follows: $a=True$, $b=True$, $c=True$, $d=True$, $e=True$, and so on and so forth for all n variables. The second truth assignment is as follows: $a=False$, $b=True$, $c=True$, $d=True$, $e=True$, and so on and so forth for all n variables. } 
\begin{itemize}
\item When n=2, there are 3 clauses in the Boolean formula in PCNF that can be satisfiable with two different truth assignments  and these clauses are as follows:
    \begin{itemize}
    \item 1 clause with a single variable: $(b)$.
    \item 2 clauses with double variables: $({a}\vee {b})$ and $(\sim {a}\vee {b})$.
    \end{itemize}
\item When n=3, there are 15 clauses in the Boolean formula in PCNF that can be satisfiable with two different truth assignments and these clauses are as follows:
    \begin{itemize}
    \item 2 clauses with a single variable:  $(b)$ and $(c)$.
    \item 7 clauses with double variables: $({a}\vee {b})$,  $(\sim {a}\vee {b})$,  $({a}\vee {c})$, $(\sim {a}\vee {c})$,  $({b}\vee {c})$, $({b}\vee \sim {c})$ and $(\sim {b}\vee {c})$
    \item 6 clause with triple variables: $({a}\vee {b} \vee {c})$, $({a}\vee {b} \vee \sim {c})$, $({a}\vee \sim {b} \vee {c})$, $(\sim {a}\vee {b} \vee {c})$, $(\sim {a}\vee {b} \vee \sim {c})$ and $( \sim {a}\vee \sim {b} \vee {c})$.
    \end{itemize}
\item When n=4, there are 57 clauses in the Boolean formula in PCNF formula that can be satisfiable with two different truth assignments and these clauses are as follows:
    \begin{itemize}
    \item 3 clauses with a single variable:  $(b)$, $(c)$ and $(d)$.
    \item 15 clauses with double variables: $({a}\vee {b})$, $(\sim {a}\vee {b})$, $({a}\vee {c})$, $(\sim {a}\vee {c})$, $({a}\vee {d})$, $(\sim {a}\vee {d})$, $({b}\vee {c})$, $({b}\vee \sim {c})$, $(\sim {b}\vee {c})$, $({b}\vee {d})$, $({b}\vee \sim {d})$, $(\sim {b}\vee {d})$, $({c}\vee {d})$, $({c}\vee \sim {d})$ and $(\sim {c}\vee {d})$.
    \item 25 clauses with triple variables: 
$({a}\vee {b} \vee {c})$, $({a}\vee {b} \vee \sim {c})$, $({a}\vee \sim {b} \vee {c})$, 
$(\sim {a}\vee {b} \vee {c})$, $(\sim {a}\vee {b} \vee \sim {c})$, $( \sim {a}\vee \sim {b} \vee {c})$, 
$({a}\vee {b} \vee {d})$, $({a}\vee {b} \vee \sim {d})$, $({a}\vee \sim {b} \vee {d})$, $(\sim {a}\vee {b} \vee {d})$, $(\sim {a}\vee {b} \vee \sim {d})$, $( \sim {a}\vee \sim {b} \vee {d})$, 
$({a}\vee {c} \vee {d})$, $({a}\vee {c} \vee \sim {d})$, $({a}\vee \sim {c} \vee {d})$, $(\sim {a}\vee {c} \vee {d})$, $(\sim {a}\vee {c} \vee \sim {d})$, $( \sim {a}\vee \sim {c} \vee {d})$,
$({b}\vee {c} \vee {d})$, $({b}\vee {c} \vee \sim {d})$, $({b}\vee \sim {c} \vee {d})$, $({b}\vee \sim {c} \vee \sim {d})$, $(\sim {b}\vee {c} \vee {d})$, $(\sim {b}\vee {c} \vee \sim  {d})$ and $(\sim {b}\vee \sim {c} \vee {d})$.
   \item 14 clauses with quadruple variables:
$({a}\vee {b} \vee {c} \vee {d})$, $({a}\vee {b} \vee {c} \vee \sim {d})$, 
$({a}\vee {b} \vee \sim {c} \vee {d})$, $({a}\vee {b} \vee \sim {c} \vee \sim {d})$, 
$({a}\vee \sim{b} \vee {c} \vee {d})$, $({a}\vee \sim{b} \vee {c} \vee \sim {d})$, 
$({a}\vee \sim{b} \vee \sim {c} \vee {d})$, $(\sim {a}\vee {b} \vee {c} \vee {d})$, 
$(\sim {a}\vee {b} \vee {c} \vee \sim {d})$, $(\sim {a}\vee {b} \vee \sim {c} \vee {d})$, 
$(\sim {a}\vee {b} \vee \sim {c} \vee \sim {d})$, $(\sim {a}\vee \sim{b} \vee {c} \vee {d})$, 
$( \sim{a}\vee \sim{b} \vee {c} \vee \sim {d})$ and $(\sim {a}\vee \sim{b} \vee \sim {c} \vee {d})$.
    \end{itemize}
\end{itemize}

When n=5 then there are 195 clauses in the Boolean formula in PCNF that can be satisfiable with two different truth assignments: 4 clauses with a single variable, 26 clauses with double variables, 64 clauses with triple variables, 71 clauses with quadruple variables and 30 clauses with quintuple variables. When n=6 then there are 633 clauses in the Boolean formula in PCNF that can be satisfiable with two different truth assignments: 5 clauses with a single variable, 40 clauses with double variables, 130 clauses with triple variables, 215 clauses with quadruple variables, 181 clauses with quintuple variables and 62 clauses with sextuple variables. 

We only need to remove one clause with a single variable and this can be expressed with the following formula: $s_{1}(n) = \binom{n-1} {0}$. We then would always remove one additional clause with a double variable. This can be expressed with the following formula: $s_{2}(n) = \binom{n-1} {1}$. The number of clauses with triple variables that we would need to remove can also be expressed with a similar formula: $s_{3}(n) = \binom{n-1} {2}$. And the same thing can be said about the clauses with quadruple variables, where the formula is $s_{4}(n) = \binom{n-1} {3}$. And so on and so forth as n increases. Once again, the binomial values in each $s_{k}$ correspond to the values at the k$^{th}$ diagonal of the Pascal's triangle starting from the right and going left as k increases, where k=0 is right most diagonal with all 1 values. The combined formula that gives the total number of clauses we would need to remove to achieve a doubly satisfiable formula from a Boolean formula in PCNF with f(n) clauses that is already satisfiable is $s(n) = \sum_{i=0}^{n-1} \binom{n-1} {i}$.

Let's assume that s(n) is correct for n. We now look at n\`{}= n+1, the successor of n. It can be seen that s(n\`{}) properly determines the total number of clauses we would need to remove to achieve a doubly satisfiable formula when there are n+1 variables in a Boolean formula in PCNF that is already satisfiable. Thus $s(n+1) = s(n) + 2^{n-1} $, where $2^{n-1}$ equals to the difference in the values between each row of Pascal's triangle. 

Therefore, if a Boolean formula in Precise Conjunctive Normal Form has more than g(n) = m(n) - r(n) - s(n) = $\sum_{i=1}^{n} 2^{i} \binom{n} {i} - \sum_{i=1}^{n} \binom{n} {i} - \sum_{i=0}^{n-1} \binom{n-1} {i} $  clauses then it has either a unique satisfying truth assignment or no satisfying truth assignment. $\blacksquare$

\end{proof}
\end{theorem}

The formula $ g(n) =  \sum_{i=1}^{n} 2^{i} \binom{n} {i} - \sum_{i=1}^{n} \binom{n} {i} -  \sum_{i=0}^{n-1} \binom{n-1} {i} $ actually simplifies to $3^{n}-2^{n}-2^{n-1}$. This is the final point of double satisfiability. 

\begin{definition}\normalfont

The promise in the ${Unambiguous-SAT}$ problem is that the given Boolean formula in CNF is guaranteed to have a unique satisfying truth assignment or none. If we restrict to only allow Boolean formulas in PCNF, then the interval $g(n)<M \leq f(n)$ is the \emph{natural range} of ${Unambiguous-SAT}$, because all of the formulas with M clauses have either a unique satisfying truth assignment or none. 

\end{definition}

\section{Solving PCNF formulas in the \emph{natural range} of ${Unambiguous-SAT}$} 

In this section, we present several corollaries derived from the theorems in the previous section. The functions in these corollaries determine the unsatisfiability of some Boolean formulas in PCNF. We also provide an algorithm that can establish the unsatisfiability of some Boolean formulas in PCNF. 

\subsection{Functions that count literals and variables to determine unsatisfiability}
\begin{corollary}\normalfont
If a Boolean formula in Precise Conjunctive Normal Form has a variable (both in complemented and uncomplemented forms) that appears more than $v(n) = 3^{n-1}+3^{n-1}-2^{n-1}$ times then it has no satisfying truth assignment.

\begin{proof}\normalfont
Immediate consequence of Theorem 2. We can select one variable and only keep the clauses that contain that variable (both in complemented and uncomplemented forms) from the Boolean formulas in PCNF in Theorem 2. This yields the maximum number of times a variable can appear (both in complemented and uncomplemented forms) in a Boolean formula in PCNF while that formula is still satisfiable. If that variable appears one more time (either in complemented or uncomplemented form) then that Boolean formula in PCNF has no satisfying truth assignment. $\blacksquare$

\end{proof}
\end{corollary}

In essence, the occurrence of a variable is tied to the occurrence of a clause since we do not allow repetition of clauses in a formula nor repetition of variables in a clause for Boolean formulas in PCNF.

\newpage

\begin{corollary}\normalfont
The maximum number of times a literal (a variable or its complement) can appear in a Boolean formula in Precise Conjunctive Normal Form is  $p(n) = 3^{n-1}$ times. 
\begin{proof}\normalfont
Immediate consequence of Theorem 1. We can select one literal and only keep the clauses that contain that literal from the Boolean formulas in PCNF in Theorem 1. This yields the maximum number of times a literal (either variable or its complement) can appear in a Boolean formula in PCNF. $\blacksquare$
\end{proof}
\end{corollary}

\begin{corollary}\normalfont
If a Boolean formula in Precise Conjunctive Normal Form has a variable that appears exactly $p(n) = 3^{n-1}$ times and its complement appears more than $q(n) = 3^{n-1}-2^{n-1}$ times then it has no satisfying truth assignment. 
\begin{proof}\normalfont
Immediate consequence of Corollary 1 and Corollary 2. We know from Corollary 2 the maximum number of times a literal (a variable or its complement) can appear, which is $p(n)$, in a Boolean formula in PCNF. We also know from Corollary 1 that if a variable (both in complemented and uncomplemented forms) appears more than $v(n)$ times then that Boolean formula in PCNF is not satisfiable. Therefore, $q(n)$ is the difference of $v(n)$ - $p(n)$. We can also simply derive this from Theorem 2. $\blacksquare$
\end{proof}
\end{corollary}

It is clear that $v(n) < f(n)$,  $p(n) < f(n)$ and $q(n) < f(n)$. Accordingly, we can use these functions to determine the unsatisfiability of some Boolean formulas in PCNF in the \emph{natural range} of ${Unambiguous-SAT}$ and beyond. 

\subsection{Algorithm that counts related clauses to determine unsatisfiability}

The Algorithm that we present makes one scan of the input Boolean formula in PCNF. During this scan, for each clause, it removes all of the negation operators from its variables and creates a dictionary with the resulting variables. We'll call this dictionary $C$, where the $key$ is the variables of the clauses in lexicographical order without their negation operators and the $value$ is the total number of times it has seen such variable combination. It also creates yet another dictionary. We'll call this dictionary $U$, where the $key$ is the number of variables in the clause it just examined and the $value$ is the maximum number of times that variable combination can occur in Boolean formulas in PCNF.\footnote{This value is $2^{n}$ and can be deduced from Theorem 1.} Of course, there could be numerous clauses with that many variables, so it will first check to see if that $key$ already exists in $U$ before creating a new $key$ and $value$ pair. It then goes through dictionary $C$ and checks to see if the $value$ of any $key$ matches the $value$ of the corresponding $key$ in $U$. If so, then the given Boolean formula in PCNF is not satisfiable. It is easy to see why such type of formulas can not be satisfiable.\footnote{This can be deduced from Theorem 2.}

\begin{algorithm}[H]
\caption{An algorithm that determines the unsatisfiability for some Boolean formulas in PCNF}
 \hspace*{\algorithmicindent} \textbf{Input} $L$ {$\triangleright$ List of clauses in PCNF} \\
 \hspace*{\algorithmicindent} \textbf{Output} $[``Unsatisfiable"\, ,``Do\,not\, know\,satisfiability"]$
\begin{algorithmic}
\State $C = \{\}$ {$\triangleright$ Clauses without any $\sim$ operator and their respective count }\footnotemark{}
\State $U = \{\}$ {$\triangleright$ Unsatisfiability check for class of clauses }
\State $m \gets |L|${$\triangleright$ Total number of clauses}
\State $i \gets 0$
\While{$ i \neq m$}
    \State$c \gets re.sub('\sim','',L[i])$
     \If{$c \in C$}
         \State $C[c] \gets C[c] + 1$
     \Else
          \State $C[c] \gets 1$
     \EndIf
     \If{$|c| \notin U$}
           \State $U[|c|] \gets 2^{(|c|)}$
     \EndIf

    \State$i \gets i + 1$
\EndWhile
\For{$(key, value) \in C.iteritems()$}
    \If {$U[|key|] == value$} 
         \State $\textbf{Output}{(``Unsatisfiable")}$
    \EndIf
\EndFor
\State $\textbf{Output:}{(``Do\,not\, know\,satisfiability")}$
\end{algorithmic}
\end{algorithm}\footnotetext{We can actually create a dictionary with a list of values. Where the first value in the list would be the total number of times we have seen such variable combination and the second value in the list would be the maximum number of times that variable combination can occur. And the algorithm would terminate if the first value ever equals the second value. However, we had not come up with this version back then so we'll keep it as is!}

We'll look at several examples of how this algorithm determines unsatisfiability. If the Boolean formula in PCNF has the following 2 clauses: $(a)$ and $(\sim a)$ then it is easy to see that it would not be satisfiable. These clauses would be saved in dictionary $C$ as $key=`a\textrm'$ and $value=`2\textrm'$ and the corresponding key value pair in $U$ would be $key=`1\textrm'$ and $value=`2\textrm'$. If the Boolean formula in PCNF has the following 4 clauses: $({a}\vee {b})$, $({a}\vee \sim {b})$, $(\sim {a}\vee {b})$ and $(\sim {a}\vee \sim {b})$ then there does not exists any truth assignment to its variables that would make the formula evaluate to True. These clauses would be saved in dictionary $C$ as $key=`ab\textrm'$ and $value=`4\textrm'$ and the corresponding key value pair in $U$ would be $key=`2\textrm'$ and $value=`4\textrm'$. If the Boolean formula in PCNF has the following 8 clauses: $({a}\vee {b} \vee {c})$, $({a}\vee {b} \vee \sim {c})$, $({a}\vee \sim {b} \vee {c})$, $({a}\vee \sim {b} \vee \sim {c})$, $(\sim {a}\vee {b} \vee {c})$, $(\sim {a}\vee {b} \vee \sim {c})$, $( \sim {a}\vee \sim {b} \vee {c})$ and $( \sim {a}\vee \sim {b} \vee \sim {c})$ then none of the eight different truth assignments to its variables would make the formula evaluate to True. These clauses would be saved in dictionary $C$ as $key=`abc\textrm'$ and $value=`8\textrm'$ and the corresponding key value pair in $U$ would be $key=`3\textrm'$ and $value=`8\textrm'$. Essentially, this algorithm determines the unsatisfiability of some Boolean formulas in PCNF in the \emph{natural range} of ${Unambiguous-SAT}$ and beyond.

The functions in the previous subsection and the algorithm in this subsection can be combined to create a more thorough algorithm to determine the unsatisfiability of Boolean formulas in PCNF in the \emph{natural range} of ${Unambiguous-SAT}$. However, this would still not achieve a complete algorithm that can determine the unsatisfiabiliy of all of the Boolean formulas in PCNF in the \emph{natural range} of ${Unambiguous-SAT}$. For example, the following Boolean formula in PCNF is in the \emph{natural range} of ${Unambiguous-SAT}$ and is unsatisfiable: $({a})$$\wedge$$(\sim{b})$$\wedge$$(\sim{c})$$\wedge$ $({a}\vee {b})$$\wedge$$({a}\vee \sim {b})$$\wedge$$(\sim {a}\vee \sim {b})$$\wedge$$({a}\vee {c})$$\wedge$$({a}\vee \sim {c})$$\wedge$$(\sim {a}\vee \sim {c})$$\wedge$$({b}\vee {c})$$\wedge$$(\sim {b}\vee \sim {c})$$\wedge$ $({a}\vee {b} \vee \sim {c})$$\wedge$$({a}\vee \sim {b} \vee {c})$$\wedge$$({a}\vee \sim {b} \vee \sim {c})$$\wedge$$(\sim {a}\vee {b} \vee {c})$$\wedge$$(\sim {a}\vee {b} \vee \sim {c})$$\wedge$$( \sim {a}\vee \sim {b} \vee {c})$. However, neither the functions in the previous subsection nor the algorithm in this subsection can determine that it is unsatisfiable. The number of clauses is 17, which is within the \emph{natural range} of ${Unambiguous-SAT}$, where n=3 and $g(3)=15<17<19=f(3)$. We would need a variable to occur more than $v(3)=14$ times to establish unsatisfiability, but none of them do. The variable counts are as follows: \{a=13, b=12, c=12\}. Furthermore, we would need a literal to occur exactly $p(3)=9$ times and its complement to occur more than $q(3)=5$ times to establish unsatisfiability, but none of the literals and their complements have this presence. The literal counts are as follows: \{a=8, $\sim{a}$=5, b=5, $\sim{b}$=7, c=5, $\sim{c}$=7\}. On the other hand, the algorithm fails to catch the unsatisfiability as well because each class of clauses appears less than the required number of times for it to determine unsatisfiability. In this instance, dictionary $C$ has the following $key$ and $value$ pairs:$ \{`a\textrm':`1\textrm',`b\textrm':`1\textrm',`c\textrm':`1\textrm',`ab\textrm':`3\textrm', `ac\textrm':`3\textrm', `bc\textrm':`2\textrm', `abc\textrm':`6\textrm'\}$ and dictionary $U$ has the following $key$ and $value$ pairs:$\{`1\textrm':`2\textrm',`2\textrm':`4\textrm',`3\textrm':`8\textrm'\}$.

\section{Conclusion}

We have defined the \emph{natural range} of ${Unambiguous-SAT}$ in terms of the number of clauses for Boolean formulas in PCNF. We have then provided several ways to determine the unsatisfiability of some Boolean formulas in PCNF. However, none of these approaches, even in combination, were complete enough to determine the unsatisfiability of all of the unsatisfiable instances in the \emph{natural range} of ${Unambiguous-SAT}$. We believe that resolution-refutation\footnote{Various complete methods for solving Boolean formulas in CNF are discussed in chapter 3 of  \cite{AMH09}.} could be used to solve these edge cases that our methods cannot solve. A complete algorithm could be constructed by first applying the counting functions, then the algorithm we provided, and finally resolution-refutation. Or perhaps there exists a much easier way to determine unsatisfiability in the \emph{natural range} of ${Unambiguous-SAT}$.

However, even if we could solve all of the instances in the \emph{natural range} of ${Unambiguous-SAT}$ in polynomial time, this would not mean that {\bf RP} = {\bf NP}. This is because the resultant formula after applying the Valiant-Vazirani isolation lemma is not necessarily in the  \emph{natural range} of ${Unambiguous-SAT}$. This can be more easily seen from the example in chapter 18 of \cite{P94}, where the isolation lemma introduces new variables, thereby moving the resultant formula further away from the \emph{natural range} of ${Unambiguous-SAT}$. The only way to show that {\bf RP} = {\bf NP}, if all of the instances in the \emph{natural range} of ${Unambiguous-SAT}$ could be proven to be solvable in polynomial time, is to come up with a new isolation lemma. This new isolation lemma would need to ensure that its output is both in PCNF and in the \emph{natural range} of ${Unambiguous-SAT}$. 

\bibliographystyle{alpha}
\bibliography{A5_Natural_Range_of_Unambiguous_SAT}
\newpage
\section{Appendix}
\begin{algorithm}[H]
\caption{An algorithm that turns a CNF to PCNF}
 \hspace*{\algorithmicindent} \textbf{Input} $CNF$ {$\triangleright$ List of Clauses in CNF} \\
 \hspace*{\algorithmicindent} \textbf{Output} $PCNF$ {$\triangleright$ Set of Clauses in PCNF}
\begin{algorithmic}
\State $PCNF=()${$\triangleright$ empty set }
\For{$c \in CNF$}
    \State{$new\_c=(c)$}{$\triangleright$ remove literals that appear more than once in a clause }
    \State{$temp\_c=()$}
     \For{$l \in new\_c$}
          \State{$temp\_c.add(re.sub('\sim','',l))$}
      \EndFor
      \If{$|new\_c| = |temp\_c|$}{$\triangleright$ remove clauses that have a variable \& its complement }
          \State{$PCNF.add(new\_c)$}{$\triangleright$ remove repeated clauses }
      \EndIf
\EndFor
\end{algorithmic}
\end{algorithm}

\end{document}